\documentclass[12pt,twoside]{amsart}

\usepackage{amssymb}
\usepackage[mathscr]{eucal}

\newtheorem{prop}{Proposition}

\newtheorem{cor}{Corollary}

\theoremstyle{definition}

\newcommand{\ds}{\displaystyle}

\def\EXP{\mbox{{e}}}
\newcommand{\sltwol}{\widetilde{\mathfrak{sl}}_2}
\newcommand{\sk}[1]{\left(#1\right)}

\renewcommand{\L}{L}
\renewcommand{\u}{u}
\newcommand{\w}{w}
\newcommand{\uop}{\mathbf{u}}
\newcommand{\wop}{\mathbf{w}}

\newcommand{\la}{\lambda}

\newcommand{\ZZ}{\mathbb{Z}}
\newcommand{\CCC}{\mathbb{C}}
\newcommand{\pr}{\partial}
\renewcommand{\r}[1]{(\ref{#1})}
\newcommand{\nn}{\nonumber}

\newcommand{\al}{\alpha}
\newcommand{\be}{\beta}
\newcommand{\ka}{\kappa}
\newcommand{\Tmon}{\widehat{T}}
\newcommand{\Ttra}{T}
\newcommand{\id}{{\rm id}}
\newcommand{\ot}{\otimes}
\newcommand{\dis}[1]{\displaystyle{#1}}

\newcommand{\ep}{\varepsilon}
\newcommand{\proiz}[3]{\frac{\pr^{#1} #2}{\pr x_{#3}^{#1}}}

\newcommand{\Tau}{\mbox{\Large $\tau$}}
\newcommand{\ff}{\phi}
\newcommand{\Om}{{\Delta}}
\newcommand{\Mo}{{\Delta}}

\newcommand{\LL}{\mathcal{L}}
\newcommand{\MM}{\overline{\mathcal{L}}}
\newcommand{\WW}{\mathcal{W}}
\newcommand{\VV}{\overline{\mathcal{W}}}
\newcommand{\BB}{\mathcal{B}}
\newcommand{\CC}{\mathcal{C}}
\newcommand{\diag}{{\rm diag}}
\newcommand{\vep}{\varepsilon}
\newcommand{\ww}{\psi}
\newcommand{\vv}{\overline\psi}
\newcommand{\www}{\Psi}
\newcommand{\vvv}{\overline\Psi}
\let\ov=\overline
\newcommand{\TTau}{\Tau}
\newcommand{\TTheta}{\Theta}
\newcommand{\UUU}{{\mathsf{U}}}
\newcommand{\WWW}{{\mathsf{W}}}
\newcommand{\UUUU}{{\ov\UUU}}
\newcommand{\WWWW}{{\ov\WWW}}

\newcommand{\vph}[1]{\varphi^{(#1)}}
\newcommand{\vvph}[1]{\ov{\varphi}^{(#1)}}
\newcommand{\amp}{f}
\begin{document}

\begin{flushright}
ITEP-TH-40/01
\end{flushright}

\vspace{2cm}

\title[]{Relativistic Toda chain at root of unity III.\\
Relativistic Toda chain hierarchy}%
\author{S. Pakuliak}
\address{Bogoliubov Laboratory of Theoretical Physics, Joint Institute
for Nuclear Research, Dubna 141980, Russia and Max-Planck Institut
f\"ur Mathematik, Vivatsgasse 7, 53111 Bonn, Germany}%
\email{pakuliak@thsun1.jinr.ru}%
\author{S. Sergeev}%
\address{Bogoliubov Laboratory of Theoretical Physics, Joint Institute
for Nuclear Research, Dubna 141980, Russia}
\email{sergeev@thsun1.jinr.ru}

\thanks{This work was supported in part by the grant INTAS OPEN 00-00055.
S.P.'s work was supported by the grant RFBR 01-01-00539 and grant
for support of scientific schools RFBR 00-15-96557, S.S's work
was supported by the grant RFBR 01-01-00201.}%


\begin{abstract}
The hierarchy of the classical nonlinear integrable equations
associated with relativistic Toda chain model is considered. It is
formulated for the $N$-powers of the quantum operators of the
corresponding quantum integrable models. Following the ideas of the
paper \cite{KMZ} it is shown how one can obtain such a system from 2D
Toda lattice system. The reduction procedure is described explicitly.
The soliton solutions for the relativistic Toda chain  are
constructed using results of  \cite{ohta}
 in terms of the rational tau-functions.
The vanishing properties of these tau-functions are investigated.
\end{abstract}
\maketitle

\section*{Introduction}

This paper is a supplement to the papers \cite{SP,SP1}.
It contains proofs of several  statements used in
these papers and related to the classical relativistic Toda chain
(RTCh) model. The papers \cite{SP,SP1} deal with the quantum  integrable
model based on the Weyl algebra at the roots of unity
\begin{equation}\label{Weyl-algebra}
\ds \uop_m^{}\cdot\wop_m^{}\,=\,
\omega\,\wop_m^{}\cdot\uop_m^{}\;,
\quad
\ds\omega\;=\;\EXP^{2\,\pi\,i/ N}\;.
\end{equation}
Namely, the simplest possible model which is
the  RTCh model at the root of unity was considered there.
The parameter $\omega$ being
different from unity describes the deviation of the quantum model from
the corresponding classical counterpart.

In the quantum integrable
models of this type, in contrast to the models where the quantization
parameter is in general position, there  exists a non-standard
classical limit. In usual, in order to get a classical model from the
quantum one, the quantization parameter is sent to unity and the
Poisson bracket appears in this classical limit as normalized
commutator.
In this limit the quantum observables becomes classical ones, namely,
$\CCC$-valued functions depending on the evolution parameters (times
or coordinates).
In the quantum integrable models based on the Weyl algebra
at the root of unity there is another possibility. One may consider
the $N$-th powers of the quantum dynamical variables
\begin{equation}\label{N-powers}
\u_m\equiv \uop^N_m\;,\quad \w_m\equiv \wop^N_m\;.
\end{equation}
Writing the deformation parameter in the form
\begin{equation}\label{BS-limit}
\ds\omega\;=\;\EXP^{2\,\pi\,i/ N}\cdot\EXP^{\varepsilon/N^2}
\end{equation}
and considering the limit $\varepsilon\to 0$ \cite{BR-unpublished}
one may obtain from the commutation relations
\r{Weyl-algebra} the Poisson bracket
\begin{equation}\label{P-br}
\{\u_m,\w_n\}=\delta_{nm}\u_m\w_n\;.
\end{equation}
It is clear that logarithms of the dynamical variables
form the canonical symplectic form
\begin{equation}\label{symplectic}
\ds\Omega=\sum \; {d\,\ln\;u_m}\,\wedge\,{d\,\ln\;w_m}\;.
\end{equation}

The important feature of this unusual classical limit is that the
quantum variables do not disappear and their $N$-th powers being the
parameters of the unitary representations of the Weyl algebra form the
classical integrable system. The quantum analysis of such a combined
systems becomes more transparent. In particular, in the paper
\cite{SP1} a modified $\mathbf{Q}$-operator  was constructed for the
quantum RTCh model at the roots of unity using several result from the
corresponding classical integrable system.

The aim of this paper is to
prove some of these  results from the point of view of the standard
theory of integrable nonlinear differential equations.
In this  approach the integrable equation
appears usually as a representative of some family of equations
sometimes called  hierarchy of integrable equations. In some cases,
these hierarchies are  certain reductions of a most general
system of equations. The subject of the present paper is
the RTCh equation and the associated hierarchy.
It is known that this model
can be obtained as a  reduction of
the 2D Toda lattice (2DTL) system \cite{KMZ,ohta} and its integrable
discrete analogs.

Relativistic generalizations of the classical
integrable equations attracted much attention after seminal work
by S.~Ruijsenaars \cite{R} and have been investigated in several
papers (see, for example, \cite{BR,KT,Su}).
The bilinear formalism for one of most simple example
of these relativistic
generalizations, namely, for RTCh equation was presented in
\cite{ohta}. In the latter paper the $M$-soliton solution to this equation
were constructed in terms of Casorati determinants.
We show that these solutions coincide identically with  the standard soliton
tau-functions constructed in KP or 2DTL theories \cite{DJKM,DJM,UT}
providing that
parameters of solitons lies on the rational
curve, which is related to Baxter's curve appeared in the quantum
integrable models based on the Weyl algebra at the root of unity.

The paper is organized as follows. In Section~\ref{first} we consider
RTCh model in finite volume (with lattice size equal to $M$)
 and using $L$-operator approach construct
 RTCh hierarchy in the framework of slightly modified AKNS
construction.  In Section~\ref{2D-Toda} we obtain
the RTCh hierarchy as certain reduction from 2DTL
theory following ideas of the paper \cite{KMZ}. In last Section
the soliton solutions to the first equation of RTCh hierarchy are
considered and identified with those used in papers \cite{SP,SP1}.
The Appendix contains a short review of Casoratian technique, which
helped authors of the paper \cite{ohta} to find soliton solutions of RTCh
equation.

\section{RTCh hierarchy in $r$-matrix formalism}
\label{first}

In order to establish relation
with the papers \cite{SP,SP1} we construct in
this Section the
RTCh hierarchy using language of $L$-operators and
$r$-matrix formalism. We will show that RTCh hierarchy can be obtained
as slightly modified AKNS system.
In particular, we will explain the meaning of
the constant shift in the corresponding trigonometric classical
$r$-matrix mentioned  in the paper \cite{KT}.

The $L$-operator for RTCh in the form of $2\times2$  matrix
\begin{equation}\label{L-RTCh}
\L_m=\frac{1}{\sqrt{\u_m\w_m}}
\sk{\begin{array}{cc}z^2+\ka\u_m\w_m&z\,\u_m\\
z\,\w_m&0\end{array}}
\end{equation}
appeared first in \cite{KT}. In \r{L-RTCh} $z$ is a spectral
parameter, $\ka\in\CCC$ is the  complex
parameter and $\u_m$, $\w_m$, $m\in\ZZ$ satisfy
the local Poisson bracket \r{P-br}.
This $L$-operator\footnote{We used here a gauge-equivalent dependence
of the $L$-operator \r{L-RTCh} on the
spectral parameter in contrast to \cite{SP,SP1} and also changed its
normalization.}
is associated with  lattice site $m$. In this
section we will consider the lattice of the finite volume, so
$m=1,\ldots,M$.

It is easy to show \cite{KT} that the Poisson bracket
\r{P-br} can be rewritten in $r$-matrix form \cite{FT,KT}
\begin{equation}\label{P-b-r}
\{\L_m(z)\mathop{,}\limits^\otimes\L_n(y)\}=\delta_{nm}
[\tilde r(z,y),\L_n(z)\otimes\L_m(y)]\ ,
\end{equation}
where $\tilde r(z,y)=r(z,y)+s$ and
\begin{equation}\label{r-mat}
r(z,y)=\sk{\begin{array}{cccc}\al(z,y)&0&0&0\\
0&0&\be(z,y)&0\\ 0&\be(z,y)&0&0\\ 0&0&0&\al(z,y)
\end{array}},
\end{equation}
$$
\al(z,y)=\frac{1}{2}\,\frac{z^2+y^2}{z^2-y^2}\;,\quad
\be(z,y)=\frac{zy}{z^2-y^2}\;,
$$
is the standard trigonometric classical $r$-matrix for $\sltwol$ and
\begin{equation}\label{r-s}
s=\sk{\begin{array}{cccc}0&0&0&0\\
0&-1/2&0&0\\ 0&0&1/2&0\\ 0&0&0&0
\end{array}}
\end{equation}
is a constant shift.  One may show that
after quantization
this constant shift
is related
to twisting of XXZ quantum $R$-matrix
in the commutation relation of the corresponding quantum $L$-operators
(see \cite{BS,SP}). In the quantum case the twisting of the
quantum $R$-matrix is necessary in order to use minimal
finite-dimensional representations
of the Weyl algebra at the root of unity. On the classical level
 such a shift of the classical $r$-matrix will simplify
the corresponding classical integrable hierarchy by removing higher
non-linear terms (see discussion around \r{****}).

In order to introduce the evolution of  $L$-operator
\r{L-RTCh} with respect to continuous times one have to consider
the monodromy matrix
\begin{equation}\label{monod}
\Tmon=\L_1\;\L_2\;\ldots\;\L_M=\sk{\begin{array}{cc}
A(z^2)&z\,B(z^2)\\ z\,C(z^2)&D(z^2)\end{array}},
\end{equation}
where $A(z^2)$, $B(z^2)$, $C(z^2)$ and $D(z^2)$ are polynomials
with respect to $z^2$ of the orders $M$, $M-1$, $M-1$ and $M-2$
respectively.  The commutation relation \r{P-b-r} shows that
the coefficients $T_k$, $k=0,\ldots, M$
 of the trace of the monodromy matrix \r{monod}
\begin{equation}\label{trace}
\Ttra(z^2)=A(z^2)+D(z^2)=\sum_{k=0}^M T_k\,z^{2k}
\end{equation}
are in involution with respect to this Poisson bracket
$$
\{T_k,T_n\}=0
$$
and can be taken as  Hamiltonians which correspond to evolution
of the $L$-operator \r{L-RTCh} with respect to times $t_k$
\begin{equation}\label{evol}
\frac{\pr L_m}{\pr  t_k}=\{L_m,T_k\}\ .
\end{equation}

Our goal is to rewrite the equation \r{evol} in the zero curvature
form. To do this we will  use a method described in the book \cite{FT}.
We have
\begin{equation}\label{for1}
\{\L_m(z),\Ttra(y)\}=V_{m-1}(z,y)\,\L_m(z)-\L_m(z)\,V_m(z,y)\;,
\end{equation}
where
\begin{eqnarray}\label{for2}
V_m(z,y)&=&{\rm tr}_2 [(\id\ot\L_1(y))\cdots
(\id\ot\L_m(y))\,\tilde r(z,y)\,\times\nn\\
&&\times\ (\id\ot\L_{m+1}(y))\cdots (\id\ot\L_M(y))]
\end{eqnarray}
and ${\rm tr}_2(A\ot B)\equiv {\rm tr}(B)\cdot A $.
Moreover, since the singular part of the $r$-matrix
\r{r-mat} is proportional to the permutation operator the following
equality holds
$$
\mathop{\rm lim}\limits_{z^2\to y^2}(z^2-y^2)
[V_{m-1}(z,y)\,\L_m(z)-\L_m(z)\,V_m(z,y)]=0
$$
which shows that r.h.s. of \r{for1} is polynomial with respect
to $z$ and $y$
\begin{equation}\label{V-poly}
V_m(z,y)=\sum_{k=0}^M\tilde Q_m^{(k)}(z)y^{2k}\;,
\end{equation}
where
\begin{equation}\label{tildeQ}
\tilde Q_m^{(k)}(z)= \sum_{p=0}^{k}a_{kp}(T)
Q_m^{(p)}(z)\;,
\end{equation}
\begin{equation}\label{Q-str}
\begin{array}{rcl}
\ds Q_m^{(n)}(z)&=&
\ds \sk{\begin{array}{cc}{z^{2n}}/{2}&0\\0&-{z^{2n}}/{2}\end{array}}
+\sum_{k=1}^n z^{2(n-k)}
\sk{\begin{array}{cc}
{h_{2k}}/{2}  & z^{-1}e_{2k-1} \\
z^{-1} f_{2k-1}  & -{h_{2k}}/{2} \end{array}} \\[2mm]
&+&\ds
\varepsilon\sk{\begin{array}{cc}
 {h_{2n}}/{2}  & 0 \\
0        & -{h_{2n}}/{2} \end{array}}\;,\quad \varepsilon=-\frac{1}{2}
\end{array}
\end{equation}
and the entries of the matrix polynomials $Q_m^{(n)}(z)$
 $h_{2k}$, $e_{2k-1}$ and $f_{2k-1}$ are differential polynomials
of the functions $u_m$ and $w_{m+1}$ only. The coefficients $a_{kp}(T)$
in decomposition \r{tildeQ} are polynomial functions of the integrals of
 motions $T_k$.

We can see that the structure of $2\times2$ matrix polynomials
\r{Q-str}
looks very similar to those in the AKNS construction.
Let us introduce  new  time variables  $x_k$
such that evolution of $L$-operator \r{L-RTCh}
is given by the zero curvature equation
\begin{equation}\label{ze-cu}
\frac{\pr \L_m(z)}{\pr
x_n}+Q^{(n)}_{m-1}(z)\,\L_m(z)-\L_m(z)\,Q^{(n)}_m(z)
=0\;.
\end{equation}

In \r{ze-cu} we can choose integer $n$ to be any non-negative integer
and call the obtained hierarchy of equations as RTCh hierarchy.
Let us write explicitly several first equations of this hierarchy
(see the Section~\ref{soliton}, formula \r{RTCh-1-res},
 where linear combination of the systems
\r{RTCh-0} and \r{RTCh-1} is identified with RTCh equation):
\begin{equation}\label{RTCh-0}
-\dis{\frac{\pr \ln\u_m}{\pr x_0}}=
\dis{\frac{\pr \ln\w_m}{\pr x_0}}=1
\end{equation}
\begin{equation}\label{RTCh-1}
\begin{array}{rcl}
\dis{\frac{\pr \ln\u_m}{\pr x_1}}&=&\dis{\w_m(\ka\u_m+\u_{m-1})\;,}\\[3mm]
-\dis{\frac{\pr \ln\w_m}{\pr x_1}}&=&\dis{\u_m(\ka\w_m+\w_{m+1})\;,}
\end{array}
\end{equation}
\begin{equation}\label{RTCh-2}
\begin{array}{rcl}
-\dis{\frac{\pr \ln\u_m}{\pr
x_2}}&=&\u_m\w_m\w_{m+1}(\ka\u_{m}+\u_{m-1})\ +\\
&+&\u_{m-1}\w_{m-1}\w_{m}(\ka\u_{m-1}+\u_{m-2})+\w^2_{m}
(\ka\u_{m}+\u_{m-1})^2\;,\\
\dis{\frac{\pr \ln\w_m}{\pr
x_2}}&=&\u_{m-1}\u_{m}\w_{m}(\ka\w_{m}+\w_{m+1})\ +\\
&+&\u_{m}\u_{m+1}\w_{m+1}(\ka\w_{m+1}+\w_{m+2})+
\u^2_{m}
(\ka\w_{m}+\w_{m+1})^2\;.
\end{array}
\end{equation}
The compatibility condition between the systems
\r{RTCh-1} and \r{RTCh-2} falls into quite simple form:
\begin{equation}\label{RTCh-1-2}
\begin{array}{rcl}
\ds \proiz{\relax}{\u_m}{2}&=&\ds -\proiz{2}{\u_m}{1}
-2 \u_m\w_{m+1}\proiz{\relax}{\u_m}{1}\;, \\
\ds\proiz{\relax}{w_{m+1}}{2}&=&\ds\proiz{2}{\w_{m+1}}{1}
-2 \u_m\w_{m+1}\proiz{\relax}{\w_{m+1}}{1}\;.
\end{array}
\end{equation}

As we already mentioned,
the structure of the matrices $Q^{(n)}_m$ \r{Q-str}  shows that the whole
RTCh hierarchy can be obtained as a version of AKNS hierarchy
\cite{Ne,KMZ} with  small modification related to the shift matrix $s$
\r{r-s}.
In \r{Q-str} we decompose
the matrices $Q^{(n)}_m$ into two pieces. The first one correspond to
the $r$-matrix \r{r-mat} and to standard AKNS construction,
while the second one to the constant shift
matrix $s$ \r{r-s}.
If the parameter $\varepsilon$ in \r{Q-str} is free then the first
equation in the corresponding AKNS construction, written as the
zero-curvature condition for matrices \r{Q-str},
\begin{equation}\label{fir-sec}
\frac{\pr Q^{(1)}}{\pr x_2}-\frac{\pr Q^{(2)}}{\pr x_1}+[Q^{(2)},Q^{(1)}]=0\;,
\end{equation}
has the form  (we renamed $u_m\equiv e$ and $w_{m+1}\equiv f$ for simplicity)
\begin{equation}\label{****}
\begin{array}{rcl}
\ds\proiz{\relax}{e}{2}&=&\ds-\proiz{2}{e}{1}
+4\ep ef\proiz{\relax}{e}{1}+2(2\ep+1)\sk{e^2\proiz{\relax}{f}{1}
+(\ep+1)e^3f^2}\;,
\\[2mm]
\ds \proiz{\relax}{f}{2}&=&\ds \phantom{-}\proiz{2}{f}{1}
+4\ep ef\proiz{\relax}{f}{1}+2(2\ep+1)\sk{f^2\proiz{\relax}{e}{1}
-(\ep+1)e^2f^3}\;.
\end{array}
\end{equation}
It is seen from \r{****} that these nonlinear equations
are simplified essentially
at the choice of the parameter $\ep=-1/2$  by removing
the higher nonlinear terms. The same phenomena occurs
with all higher hierarchy equations.  We would like to repeat
once more that such a simplification is a classical analog of the
possibility to use the minimal representation of the Weyl algebra
in the corresponding quantum model.

Due to locality property ($Q^{(n)}_m$ depends only on $\u_m$ and
$\w_{m+1}$) the AKNS construction do not describe discrete time
evolution \r{ze-cu}.
As well as in the usual AKNS hierarchy the dependence on the
discrete time or the discrete  evolution \r{ze-cu} can be
reconstructed after
using  the B\"acklund-Schlezinger transformation \cite{JM}. The $L$-operator
\r{L-RTCh} plays the role of matrix which
generates  this transformation.
It is possible to develop this AKNS construction in full details
using the methods described in the book \cite{Ne},
but instead we will describe in the next section
the RTCh hierarchy as a reduction of 2DTL.

\section{RTCh as reduction of 2DTL}
\label{2D-Toda}

In this section we recall the basic facts on 2DTL
 model following the fundamental paper \cite{UT}. Then we
consider a certain reduction of this system which leads to the
RTCh hierarchy.

\subsection{2D Toda lattice} In \cite{UT} 2DTL was defined in
terms of $\ZZ\times\ZZ$ matrices $\LL$ and $\MM$ of the following form
\begin{equation}\label{2dtl-matrices}
\LL=\sum_{j\leq1}\diag[b_j(m)]\Lambda^j\;,\quad
\MM=\sum_{j\geq-1}\diag[c_j(m)]\Lambda^j\;,
\end{equation}
where $b_1(m)=1$, $c_{-1}(m)\neq0$ for any $m\in\ZZ$ and
$\Lambda_{m,k}\equiv \delta_{m+1,k}$ is a shift matrix.
In components, these infinite matrices can be written as follows
\begin{equation}\label{mat_in_comp}
\begin{array}{rcl}
\LL_{m,k}&=&\ds\delta_{m+1,k}+\sum_{j=0}^\infty
b_{-j}(m)\delta_{m-j,k}\;,\\
\MM_{m,k}&=&\ds c_{-1}(m)\delta_{m-1,k}+\sum_{j=0}^\infty
c_{j}(m)\delta_{m+j,k}\;.
\end{array}
\end{equation}

For arbitrary infinite matrix $A=\sum_{j\in\ZZ}\diag[a_j(m)]\Lambda^j$
we define its projections on the upper-diagonal (including diagonal)
$A_+$ and the strictly lower-diagonal (excluding diagonal) $A_-$
matrices by the formulas
\begin{equation}\label{projections}
A_+=\sum_{j\geq0}\diag[a_j(m)]\Lambda^j\;,\quad
A_-=\sum_{j<0}\diag[a_j(m)]\Lambda^j\;.
\end{equation}

2DTL is a system of differential equations for the functions
$b_{-j}(m)$, $c_j(m)$, $j\geq0$  and $c_{-1}(m)$ with respect to two
copies of the continuous parameters (times) $x=\{x_1,x_2,\ldots\}$ and
$y=\{y_1,y_2,\ldots\}$ of the form
\begin{equation}\label{2DTL}
\begin{array}{rclcrcl}
\ds
\pr_{x_s}\LL&=&[\BB_s,\LL]\;,&\quad&\ds\pr_{x_s}\MM&=&[\BB_s,\MM]\;,\\
\ds \pr_{y_s}\LL&=&[\CC_s,\LL]\;,&\quad&\ds\pr_{y_s}\MM&=&[\CC_s,\MM]\;,
\end{array}
\end{equation}
where $\BB_s=\sk{\LL^s}_+$ and $\CC_s=\sk{\MM^s}_-$.

The system \r{2DTL} can be obtained as the compatibility conditions of
the spectral problems
\begin{equation}\label{spectral}
\LL\;\WW(x,y)=\WW(x,y)\Lambda\;,\quad \MM\;\VV(x,y)=\VV(x,y)\Lambda^{-1}
\end{equation}
and the linear problems
\begin{equation}\label{linear}
\begin{array}{rclcrcl}
\ds\pr_{x_s}\WW(x,y)&=&\BB_s\WW(x,y)\;,&&
\ds\pr_{x_s}\VV(x,y)&=&\BB_s\VV(x,y)\;,\\[1mm]
\ds\pr_{y_s}\WW(x,y)&=&\CC_s\WW(x,y)\;,&&
\ds\pr_{y_s}\VV(x,y)&=&\CC_s\VV(x,y)\;,
\end{array}
\end{equation}
where $\WW(x,y)$ and $\VV(x,y)$ are Baker-Akhiezer functions.
Writing them in the form
$$
\begin{array}{rcl}
\WW(x,y)&=&\ds\sum_{j=0}^\infty
\diag[\ww_j(m;x,y)]\Lambda^{-j}\cdot \exp\sk{\xi(x,\Lambda)}\;,\\[1mm]
\VV(x,y)&=&\ds\sum_{j=0}^\infty
\diag[\vv_j(m;x,y)]\Lambda^{j}\cdot \exp\sk{\xi(y,\Lambda^{-1})}\;,\\[1mm]
\WW(x,y)^{-1}&=&\ds\exp\sk{-\xi(x,\Lambda)}\cdot\sum_{j=0}^\infty
\Lambda^{-j}\;\diag[\ww^*_j(m+1;x,y)]\;,\\[1mm]
\VV(x,y)^{-1}&=&\ds\exp\sk{-\xi(y,\Lambda^{-1})}\cdot\sum_{j=0}^\infty
\Lambda^{j}\;\diag[\vv^*_j(m+1;x,y)]\;,
\end{array}
$$
with $\ww_0(m;x,y)=\ww_0^*(m;x,y)=1$
and $\vv_0(m;x,y)\neq0$ for $m\in\ZZ$
and using a bilinear formalism one can prove \cite{UT}
that the
scalar generating series
\begin{equation}\label{gen-series}
\begin{array}{rcl}
\www(m;x,y;\lambda)&=&\ds\lambda^{m}\exp\xi(x,\lambda)
\sum_{j=0}^\infty\ww_j(m;x,y)\lambda^{-j}\;,\\[1mm]
\www^*(m;x,y;\lambda)&=&\ds\lambda^{-m}\exp\xi(-x,\lambda)
\sum_{j=0}^\infty\ww^*_j(m;x,y)\lambda^{-j}\;,\\[1mm]
\vvv(m;x,y;\lambda)&=&\ds\lambda^{m}\exp\xi(y,\lambda^{-1})
\sum_{j=0}^\infty\vv_j(m;x,y)\lambda^{j}\;,\\[1mm]
\vvv^*(m;x,y;\lambda)&=&\ds\lambda^{-m}\exp\xi(-y,\lambda^{-1})
\sum_{j=0}^\infty\vv^*_j(m;x,y)\lambda^{j}
\end{array}
\end{equation}
can be expressed in terms of family of tau-function $\tau_n(x,y)$,
$n\in\ZZ$
\begin{equation}\label{tau-2dtl}
\begin{array}{rcl}
\www(m;x,y;\la)&=&\ds\frac{\tau_m(x-\vep(\la^{-1}),y)}{\tau_m(x,y)}\;,\\[1mm]
\www^*(m;x,y;\la)&=&\ds\frac{\tau_m(x+\vep(\la^{-1}),y)}{\tau_m(x,y)}\;,\\[1mm]
\vvv(m;x,y;\la)&=&\ds\frac{\tau_{m+1}(x,y-\vep(\la))}{\tau_m(x,y)}\;,\\[1mm]
\vvv^*(m;x,y;\la)&=&\ds\frac{\tau_{m-1}(x,y+\vep(\la))}{\tau_m(x,y)}\;.
\end{array}
\end{equation}
In 
\r{gen-series} the function $\xi(x,\la)$
is
$$\xi(x,\la)=\sum_{s=1}^\infty x_s\la^s$$
and in \r{tau-2dtl}
the notation $\vep(\lambda)$ means the set $\{\la,\la^2/2,\la^3/3,\ldots\}$.

The formulas \r{tau-2dtl} and the spectral  problems \r{linear} allow
to express all the entries of the matrices $\LL$ and $\MM$ as certain
combinations of tau-function $\tau_n(x,y)$. In particular, the first
coefficient $c_{-1}(n)$ in the decomposition of the matrix $\MM$  has
the form
\begin{equation}\label{first_c}
c_{-1}(m;x,y)=\frac{\tau_{m+1}(x,y)\tau_{m-1}(x,y)}{\tau^2_{m}(x,y)}\;.
\end{equation}

\subsection{RTCh hierarchy as reduction from 2DTL}
Consider the following substitution for the entries of the matrices
$\LL$ and $\MM$ ($j\geq0$) \cite{KMZ}:
\begin{equation}\label{sub1}
\begin{array}{rcl}
b_{-j}(m)&=&\ds(-)^{j+1}\prod_{i=1}^j\u_{m-i}\prod_{i=0}^jw_{m-i}
(\kappa u_{m-j}+u_{m-j-1})\;,\\[1mm]
c_{j}(m)&=&\ds(\kappa)^{-j-1}\prod_{i=0}^{j-1}\u^{-1}_{m+i}
\prod_{i=0}^jw^{-1}_{m+i}
(\kappa u^{-1}_{m+j}+u^{-1}_{m+j+1})\;,\\[1mm]
c_{-1}(m)&=&\ds\frac{u_{m-1}}{u_m}\;.
\end{array}
\end{equation}
In \r{sub1} we introduced the functions $u_m(x,y)$ and $w_m(x,y)$
which depend on the continuous times $x_s$ and $y_s$ and
$\kappa\in\CCC$ is a parameter, the same as in the previous Section.

Comparing the last formula in \r{sub1} with \r{first_c} we can choose
the following parametrization of the function $\u_m(x,y)$ in terms of
tau-function
\begin{equation}\label{sub3}
\u_m(x,y)=\frac{\tau_m(x,y)}{\tau_{m+1}(x,y)}\;.
\end{equation}
We consider the similar ansatz for the variable $w_m(x,y)$
\begin{equation}\label{sub33}
\w_m(x,y)=\frac{\theta_{m+1}(x,y)}{\theta_{m}(x,y)}\;.
\end{equation}

 The formulas \r{sub3}, \r{sub33} and \r{sub1} allows to get
simple expressions  relations for the matrices $\LL$ and $\MM$
in terms of RTCh
dynamical variables $\u_m$ and $\w_m$. We have
\begin{prop}\label{lem1}
\begin{equation}\label{L-matrices}
\begin{array}{rcl}
\LL&=&\ds \frac{\TTheta}{\TTau}\left[
W\;\frac{1}{1+\Lambda^{-1}}\; U\; (\Lambda-\ka)
\right]\frac{\TTau}{\TTheta}\;,\\[3mm]
\MM&=&\ds \frac{\TTheta}{\TTau}\left[
\frac{1}{1-\ka^{-1}\Lambda}\; U^{-1}\; (1+\Lambda^{-1})\;W^{-1}
\right]\frac{\TTau}{\TTheta}\;,
\end{array}
\end{equation}
where $U$, $W$, $\TTau$ and $\TTheta$ are  diagonal matrices
\begin{equation}\label{diag}
U^{\pm1}=\diag[\u^{\pm1}_m]\;,\quad
W^{\pm1}=\diag[\w^{\pm1}_m]\;,\quad
\TTau^{\pm1}=\diag[\tau^{\pm1}_m]\;,\quad
\TTheta^{\pm1}=\diag[\theta^{\pm1}_m]
\end{equation}
and
the rational functions $\ds\frac{1}{1-x}$ in formulas \r{L-matrices}
is always understood as the series
$$
\frac{1}{1-x}=\sum_{j=0}^\infty x^j\;.
$$
\end{prop}
\begin{cor}
The matrices $\LL$ and $\MM$ which correspond to RTCh reduction satisfy
the relation
\begin{equation}\label{enjoy}
\LL\cdot\MM=\MM\cdot\LL=-\ka\;.
\end{equation}
\end{cor}
\proof
Substituting formulas \r{sub1} into \r{mat_in_comp} we obtain
\begin{equation}\label{calc1}
\begin{array}{rcl}
1-\Lambda^{-1}\LL&=&\ds(1+\ka) \frac{\TTheta}{\TTau_-}\;\frac{\Lambda^{-1}}{1+
\Lambda^{-1}}\;\frac{\widetilde\TTheta}{\TTau_+}\;,\\[3mm]
\MM\Lambda+\ka=&=&\ds(1+\ka) \frac{\TTheta}{\TTau}\;\frac{\ka}{\ka-
\Lambda}\;\frac{\widetilde\TTheta}{\TTau}\;,
\end{array}
\end{equation}
where we introduced the diagonal matrices
\begin{equation}\label{tilde-theta}
\TTau_\pm=\diag[\tau_{n\pm1}]\;,\quad\widetilde\TTheta=\diag[\tilde\theta_m]\;
,\quad \tilde\theta_m=
\frac{\tau_{m+1}\tau_{m-1}+\ka\tau^2_m}{(1+\ka)\theta_m}\;.
\end{equation}
Excluding from formulas \r{calc1} the diagonal matrix
$\widetilde\TTheta$ we can find first the relation  \r{enjoy}
and using this relation in \r{calc1}
we prove the statement of the Proposition.
\endproof

At the first sight the formulas \r{L-matrices} does not depend just on
variables $u_n$ and $w_n$, since they includes explicitly the diagonal
matrices $\TTau$ and $\TTheta$. But this is not true since,
we have obvious formulas
\begin{equation}\label{obvious}
\begin{array}{rcl}
\ds \frac{\TTheta}{\TTau}\;\Lambda^j\;\frac{\TTau}{\TTheta}&=&
\ds \diag\left[\prod_{i=0}^{j-1}\u^{-1}_{m+i}\w^{-1}_{m+i}\right]\;,
\quad j>0\;,\\[3mm]
\ds \frac{\TTheta}{\TTau}\;\Lambda^j\;\frac{\TTau}{\TTheta}&=&
\ds \diag\left[\prod_{i=1}^{-j}\u_{m-i}\w_{m-i}\right]\;,
\quad j<0\;.
\end{array}
\end{equation}

Using formulas \r{L-matrices} we can obtain explicit formulas for the
elements of the matrices $\BB_s$ and $\CC_s$ which describe the
evolutions of the matrices $\LL$ and $\MM$ by the formulas
\r{2DTL}. This will give the evolution equations of the functions
$u_m(x,y)$ and $w_m(x,y)$ with respect to continuous times $x_s$ and
$y_s$. Form of these equations is given by the following

\begin{prop}\label{pr-reduc}
The substitution \r{sub1} describes the reduction of 2DTL to
the couple of
the RTCh
hierarchies with respect to the sets of the times $x_s$ and $y_s$.
These hierarchies are given by the formulas
\begin{equation}\label{evol-x}
\begin{array}{rcl}
\ds\frac{\pr\ln\u_m}{\pr x_s}&=&\ds -\sum_{j_k\geq 0\atop
j_1+\ldots+j_s=s}\prod_{i=0}^{s-1}\UUU_{j_k}\sk{m+i-\sum_{r=1}^i j_r}\;,\\[3mm]
\ds\frac{\pr\ln\w_m}{\pr x_s}&=&\ds \sum_{j_k\geq 0\atop
j_1+\ldots+j_s=s}\prod_{i=0}^{s-1}\WWW_{j_k}\sk{m-i+\sum_{r=1}^i j_r}\;,
\end{array}
\end{equation}
and
\begin{equation}\label{evol-y}
\begin{array}{rcl}
\ds(-\ka)^{s}\frac{\pr\ln\u^{-1}_m}
{\pr y_s}&=&\ds \sum_{j_k\geq 0\atop
j_1+\ldots+j_s=s}\prod_{i=0}^{s-1}\UUUU_{j_k}\sk{m-i+\sum_{r=1}^i j_r}\;
,\\[3mm]
\ds(-\ka)^s\frac{\pr\ln\w^{-1}_m}
{\pr y_s}&=&-\ds \sum_{j_k\geq 0\atop
j_1+\ldots+j_s=s}\prod_{i=0}^{s-1}\WWWW_{j_k}\sk{m+i-\sum_{r=1}^i j_r}\;,
\end{array}
\end{equation}
where
\begin{equation}
\begin{array}{rcl}
\UUU_j(m)&=&(-)^j\w_m
\sk{\ka(1-\delta_{j,0})\u_{m-j+1}+\u_{m-j}}\;,\\[2mm]
\WWW_j(m)&=&(-)^j\u_m
\sk{\ka(1-\delta_{j,0})\w_{m+j-1}+\w_{m+j}}\;,
\end{array}
\end{equation}
and
\begin{equation}
\begin{array}{rcl}
\UUUU_j(m)&=&(-)^j\w^{-1}_m
\sk{\ka(1-\delta_{j,0})\u^{-1}_{m+j-1}+\u^{-1}_{m+j}}\;,\\[2mm]
\WWWW_j(m)&=&(-)^j\u^{-1}_m
\sk{\ka(1-\delta_{j,0})\w^{-1}_{m-j+1}+\w^{-1}_{m-j}}\;.
\end{array}
\end{equation}
\end{prop}

The meaning of the Proposition \ref{pr-reduc} can be understood
 as follows. Take
the equations from the system \r{2DTL}, which describe the  evolution of
the operator $\LL$ and $\MM$ with respect to the time $x_s$. This set
contains infinite number of equations, which describe the
evolution of the entries $b_{-j}(m)$  and $c_j(m)$ with respect to this time.
The statement of the Proposition \ref{pr-reduc} means that the whole
set of these equations reduces after substitution \r{sub1}
just to pair of equations given by the
formulas \r{evol-x}. The same is true for the evolutions with respect
to times $y_s$. One may  verify that first two pair of equations
from \r{evol-x} which correspond to $s=1,2$ coincide precisely with
 equations \r{RTCh-1} and \r{RTCh-2} obtained from $L$-operator approach.
A general proof
that  the hierarchy \r{ze-cu} obtained in
 $L$-operator approach coincide with hierarchy \r{evol-x} for any
 $s$ should be a repetition of the analogous proof for the equivalence
 of such a presentations in case of standard AKNS hierarchy. We do not
 go into these details here.

Explicit formulas \r{evol-x} and \r{evol-y} yield the following
\begin{cor}
\label{dis-sym}
The systems \r{evol-x} and \r{evol-y} transforms to each other after
the simultaneous transformations
\begin{equation}\label{trans}
x_s \leftarrow\!\!\rightarrow -(-\kappa)^{-s}y_s \;,\quad
u_n \leftarrow\!\!\rightarrow w_n^{-1}\;.
\end{equation}
The first equation in \r{evol-x} goes into the second equation of
\r{evol-y} while the second of \r{evol-x} to first of \r{evol-y}.
\end{cor}
The transformation \r{trans} allows to
forget about
dependence of our dynamical variables on times $y_s$ and consider only
the set of equations \r{evol-x}.

This symmetry of the reduced 2DTL is related to certain
B\"acklund transformations
of 2DTL investigated  in \cite{HS} and used in the
paper \cite{ohta} in order to obtain the bilinear formulation of RTCh
equation. Ended,  formulas \r{calc1} contains `new' tau-function
$\tilde\theta_m$ given by \r{tilde-theta} as a certain
 rational combination of $\tau_m$ and $\theta_m$.
It is actually unclear from our presentation
whether the functions $\theta_m$ and
$\tilde\theta_m$ are also the  tau-functions. The last
equation of the substitution \r{sub1} and \r{first_c} allows to
identify only $\tau_m$ with some tau-function of 2DTL.

Nevertheless, this assertion is true and
one can use the following argument to see that
the function
$\theta_m$ is also a tau-function.
It is clear that the symmetry transformation \r{trans} moves the
substitution \r{sub1} into another  one:
\begin{equation}\label{sub2}
\begin{array}{rcl}
b_{-j}(m)&=&\ds(-)^{j+1}\prod_{i=1}^j\w^{-1}_{m-i}\prod_{i=0}^j\u^{-1}_{m-i}
(\kappa \w^{-1}_{m-j}+\w^{-1}_{m-j-1})\;,\\[1mm]
c_{j}(m)&=&\ds(\kappa)^{-j-1}\prod_{i=0}^{j-1}\w_{m+i}
\prod_{i=0}^j\u_{m+i}
(\kappa \w_{m+j}+\w_{m+j+1})\;,\\[1mm]
c_{-1}(m)&=&\ds\frac{\w_{m}}{\w_{m-1}}\;,
\end{array}
\end{equation}
where now the ratio of functions $w_m$
should be identified  with
a ratio of tau-functions according to the
formulas \r{tau-2dtl}. The ansatz \r{sub33} allows to identify the
function $\theta_m$ with some another tau-function of 2DTL. This
tau-function should be different since otherwise the substitutions
\r{sub1} and \r{sub2} becomes trivial (the product $u_mw_m=1$ for any
$m\in\ZZ$ in this case).

On the other hand
if one uses the substitution \r{sub2}
in order  to get expressions for the matrices $\LL$
and $\MM$ similar to \r{calc1} a new rational combination of
tau-functions
\begin{equation}\label{tilde-tau}
\tilde\tau_m=\frac{\theta_{m+1}\theta_{m-1}+\ka\theta_m^2}{(1+\ka)\tau_m}
\end{equation}
will appear. Let us denote all these  tau-functions as a set
$\tau_m^{(k)}$
$$
\tilde\theta_m\equiv\tau^{(1)}_m\;,\quad
\tau_m\equiv\tau^{(0)}_m\;,\quad
\theta_m\equiv\tau^{(-1)}_m\;,\quad
\tilde\tau_m\equiv\tau^{(-2)}\;.
$$
We can extend the value of the index $k$ to be any integer number by
repeating the transformations given by the formulas \r{tilde-theta}
and \r{tilde-tau}. It was shown in \cite{ohta} that the whole set of
these tau-functions satisfy a reduced discrete 2DTL (the transformations
\r{tilde-theta} and \r{tilde-tau} describe the evolution with respect
to discrete parameter $k$)
 and all of them
are some tau-functions of original 2DTL hierarchy.

We will see in the
next Section that in the soliton sectors the tau-functions
$\tau_m^{(k)}$ differs by
the proportional change of the soliton amplitudes (see \r{oh6}).
In terms of the set $\tau_m^{(k)}$ the transformation \r{trans} can be
written in the form
\begin{equation}\label{trans-new}
\tau_m^{(k)} \leftarrow\!\!\rightarrow \tau_m^{(1-k)}\;,\quad
k\in\ZZ\;.
\end{equation}

\section{Soliton solutions of the RTCh equation}
\label{soliton}

In this Section we will give the  solitonic solution of the RTCh
equation obtained in \cite{ohta} in terms of rational tau-functions.
We will identify them with tau-functions
used in the papers \cite{SP,SP1} as well as with soliton tau-functions
appeared in KP \cite{DJKM,DJM} and in 2DTL \cite{UT} theories.
In order to do this we
have to transform the RTCh equation \r{RTCh-1} into bilinear form
analogous to one used in the \cite{ohta}.
The authors of this paper used
so called Casoratian technique to find a solitonic solutions to the
 RTCh equation in form of the Casorati determinants.
For the readers convenience we will describe part of this
technique relevant for our paper in the Appendix.

We start with rewriting the equations \r{RTCh-1} into bilinear form.
Actually, in order to use results on the reduction described in the
previous section, we have to consider certain linear combinations of
equations from the RTCh hierarchy. Instead of \r{RTCh-1} we consider a
linear combination of \r{RTCh-0} and \r{RTCh-1}
\begin{equation}\label{RTCh-1-res}
\begin{array}{rcl}
\dis{\frac{\pr \ln\u_m}{\pr x_1}}&=&\dis{\w_m(\ka\u_m+
\u_{m-1})-(\ka+1)\;,}\\[3mm]
-\dis{\frac{\pr \ln\w_m}{\pr
x_1}}&=&\dis{\u_m(\ka\w_m+\w_{m+1})
-(\ka+1)\;.}
\end{array}
\end{equation}
such that the system \r{RTCh-1-res} has a constant
($\u_m=\w_m^{-1}={\rm const}$) solution.

Excluding from \r{RTCh-1-res} $\w_m$ and replacing
$\u_m=\EXP^{q_m((\sqrt{1+\ka})x_1)}$ we obtain ($q'_m=\pr q_m/\pr x_1$)
\begin{equation}\label{RTCh}
q''_m=\dis{\frac{(\sqrt{1+\ka}+q'_{m-1})(\sqrt{1+\ka}+q'_{m})}
{\dis{1+\ka\, \EXP^{q_m-q_{m-1}}}}}-
\dis{\frac{(\sqrt{1+\ka}+q'_{m})(\sqrt{1+\ka}+q'_{m+1})}
{\dis{1+\ka\, \EXP^{q_{m+1}-q_{m}}}}}\;.
\end{equation}
The latter is the relativistic Toda chain equation.
In the non-relativistic limit $\ka\to\infty$ we obtain from
\r{RTCh} the standard Toda chain equation
$$
q''_m=\EXP^{q_{m-1}-q_m}-\EXP^{q_m-q_{m+1}}\;.
$$

\begin{prop} \cite{ohta}
The substitution
\begin{equation}\label{multi-soliton}
\begin{array}{rcl}
\u_m(x_1)&=&\ds\frac{\tau_{m}(x_1)}{\tau_{m+1}(x_1)}\;,\\[3mm]
\w_m(x_1)&=&\ds\frac{\theta_{m+1}(x_1)}{\theta_{m}(x_1)}\;,
\end{array}
\end{equation}
transform the equations \r{RTCh-1-res} into bilinear form
\begin{equation}\label{ohta-version}
\begin{array}{rcl}
\ds D_1^2(\tau_m\circ\tau_m)&=&\ds
2(1+\ka)(\theta_{m+1}\tilde\theta_{m-1}-\tau_m^2)\;, \\[2mm]
\ds D_1(\tau_m\circ\tau_{m-1})&=&\ds
(1+\ka)(\tau_m\tau_{m-1}-\theta_m\tilde\theta_{m-1})\;, \\[2mm]
(1+\ka)\theta_m\tilde\theta_m&=&\tau_{m+1}\tau_{m-1}+\ka\tau^2_m\;.
\end{array}
\end{equation}
\end{prop}
In \r{ohta-version} the Hirota operator $D(f\circ
g)=f'g-fg'$ have been used.

We can also find a bilinear relation between tau-functions $\tau_m$
and $\theta_m$
\begin{equation}\label{relation}
D_1(\theta_{m-1}\circ\tau_m)=(1+\ka)(\theta_m\tau_{m-1}-\theta_{m-1}\tau_{m})
\end{equation}
which is a consequence of the RTCh equation and B\"acklund transformation
$\tau_m\to\theta_m$ discussed at the end of the previous section.
In the Appendix we will show how this bilinear equation can be solved
using Casoratian technique.

Let $\Om^*$ and $\Mo$ are two independent
complex parameters which belong to the
rational curve
\begin{equation}\label{curve}
\ka=\Om^*\Mo-\Om^*-\Mo\;.
\end{equation}
In the paper \cite{SP} we used the following uniformization of the
curve \r{curve}
\begin{equation}\label{uniformization}
\begin{array}{rcl}
\Om^*_\ff&=&\EXP^{-i\ff}\sk{\sqrt{\cos^2\ff+\ka}+\cos\ff}\;,\\[3mm]
\Mo_\ff&=&\EXP^{i\ff}\sk{\sqrt{\cos^2\ff+\ka}+\cos\ff}\;
\end{array}
\end{equation}
which is convenient for restricting the model to the chain of the
finite size with periodic boundary conditions. For real $\kappa$ and
$\phi$ the parameters $\Om^*$ and $\Mo$ are complex conjugated.

Define functions
\begin{equation}\label{oh1}
\begin{array}{rcl}
\ds\vph{k}_m(x_1;\ff,\amp)&=&\ds
\frac{(1-\Om^*_\ff)^m(1-
{\Mo_\ff}^{-1})^k\EXP^{(1-\Om^*_\ff)x_1}}{(1+\ka)^{m/2}}-\\[2mm]
&-&\ds
\amp\;\frac{(1-\Mo_\ff)^m(1-{\Om^*_\ff}^{-1})^k\EXP^{(1-
\Mo_\ff)x_1}}{(1+\ka)^{m/2}}
\end{array}
\end{equation}
which depend on discrete time $m$, first continuous time  $x_1$  and some
additional integer parameter $k$. Define also the $M$-soliton tau-function
by the determinant\footnote{The superscript $^{(\nu)}$ counts the
number of B\"acklund transformations \r{relation}, not the number of
solitons as in the papers \cite{SP,SP1}.}
\begin{equation}\label{oh2}
\tau^{(\nu)}_m\left(x_1;\left.\{\ff_k,\amp_k\}\right|_{k=1}^{M}\right)=
\det\left|\vph{{m+\nu}}_{m+i-1}(x_1;\ff_j,\amp_j)\right|_{i,j=1}^M
\end{equation}

One of the statements of the paper \cite{ohta} is the following
\begin{prop}\label{prop-ohta}\cite{ohta}
The tau-functions which solve the equations \r{ohta-version}
are given by the following specialization
of the function\r{oh2}
\begin{equation}\label{oh3}
\begin{array}{rcl}
\ds\tau_m\left(x_1;\left.\{\ff_k,\amp_k\}\right|_{k=1}^{M}\right)&=&
\tau^{(0)}_m\left(x_1;\left.\{\ff_k,\amp_k\}\right|_{k=1}^{M}\right)\;,\\[2mm]
\ds\theta_m\left(x_1;\left.\{\ff_k,\amp_k\}\right|_{k=1}^{M}\right)&=&
\tau^{(-1)}_m\left(x_1;\left.\{\ff_k,\amp_k\}\right|_{k=1}^{M}\right)\;,\\[2mm]
\ds\tilde\theta_m\left(x_1;\left.\{\ff_k,\amp_k\}\right|_{k=1}^{M}\right)&=&
\tau^{(1)}_m\left(x_1;\left.\{\ff_k,\amp_k\}\right|_{k=1}^{M}\right)\;.
\end{array}
\end{equation}
\end{prop}

The tau-functions given by the
Proposition~\ref{prop-ohta} coincide with those used in the papers
\cite{SP,SP1} up to certain normalization factor. Let us change this
normalization in order to fix precise relation to formulas of these
papers and also to demonstrate the connection of these tau-functions
with standard solitonic tau-functions known from the 2DTL theory
\cite{UT}. We obtain ($\Om^*_i\equiv\Om^*_{\ff_i}$ and
$\Mo_i\equiv\Mo_{\ff_i}$)
\begin{equation}\label{oh4}
\tau^{(\nu)}_m=\prod_{i<j}(\Om^*_i-\Om^*_j)
\prod_i\frac{(-\Mo_i)^{\nu-m}}{(1-\Mo_i)^\nu}
\EXP^{x_1\sum(1-\Om^*_i)}\Tau_m^{(\nu)}\;,
\end{equation}
where
\begin{equation}\label{oh5}
\Tau_m^{(\nu)}=\prod_{i<j}(\Om^*_i-\Om^*_j)^{-1}\det
\left|(1-\Om^*_j)^{i-1}-X^{(\nu)}_m(x_1;\ff_j,\amp_j)(1-\Mo_j)^{i-1}
\frac{\Om^*_j}{\Mo_j}
\right|_{i,j=1}^M
\end{equation}
and
\begin{equation}\label{oh6}
X^{(\nu)}_m(x_1;\ff,\amp)=f\EXP^{(\Om_\ff^*-\Mo_\ff)x_1}
\sk{\frac{\Mo_\ff}{\Om_\ff^*}}^{m+1}
\sk{c(\ff)}^{-\nu}\;,\quad c(\ff)=\frac{\Om^*_\ff}{\Mo_\ff}\;
\frac{(1-\Mo_\ff)}{(1-\Om^*_\ff)}\;.
\end{equation}
It is clear that factor before $\Tau_m^{(\nu)}$
in formula \r{oh4} cannot spoil the form of
the equations \r{ohta-version} so can be dropped. Now the
tau-functions $\Tau_m\equiv\Tau^{(0)}_m$ and
$\Theta_m\equiv\Tau^{(-1)}_m$ coincide identically with those used in
papers \cite{SP,SP1}. The only difference is the shift in amplitude
\r{oh6}, but this is because we used $\u_m=\tau_m/\tau_{m+1}$
instead of $\u_m=\tau_{m-1}/\tau_{m}$ in \cite{SP,SP1}.

The formula \r{oh5} may be rewritten in the form
 analogous  to the standard presentation of solitonic tau-functions
 appeared in KP  \cite{DJKM} or 2DTL theories \cite{UT} with all
 higher continuous times but $x_1$ are freezed:
\begin{equation}\label{oh7}
\Tau_n^{(\nu)}=
\det
\left|\delta_{ij}-
X^{(\nu)}_m(x_1;\ff_j,\amp_j)\prod_{k\neq j} s_{j,k}\frac{\Om^*_i-\Mo_i}
{\Om^*_i-\Mo_j}
\right|_{i,j=1}^M
\end{equation}
where
\begin{equation}\label{s-function}
s_{j,k}=s(\ff,\ff')\;,\quad
s(\ff,\ff')=\frac{\Om^*_\ff}{\Mo_\ff}\;
\frac{\Mo_\ff-\Om^*_{\ff'}}{\Om^*_\ff-\Om^*_{\ff'}}
\end{equation}
In order to come from formula \r{oh5} to \r{oh7} one have to use
explicitly the inverse Vandermonde matrix and this is a reason
of getting product of factors $s_{j,k}$ in soliton amplitudes.
Actually, the determinant presentation \r{oh7} is equivalent to
the recurrent relation which relates the tau-functions with different
number of solitons:
$$
\Tau_m^{(\nu)}(\{\ff_k,f_k\}_{k=1}^M)=
\Tau_m^{(\nu)}(\{\ff_k,f_k\}_{k=1}^{M-1})
-X_m^{(\nu)}(\ff_M,f_M)\prod_{k=1}^{M-1}s_{M,k}
\Tau_m^{(\nu)}(\{\ff_k,d_{k,n}f_k\}_{k=1}^{M-1})\;,
$$
where the phase shift  is given by cross-ratio
\begin{equation}\label{phase}
d_{i,j}=
\frac{(\Om^*_i-\Om^*_j)(\Mo_i-\Mo_j)}
{(\Om^*_i-\Mo_j)(\Mo_i-\Om^*_j)}\ .
\end{equation}

Specializing the $M$-soliton
tau-function $\Tau^{(0)}_m(x_1;\{\ff_k,\amp_k\}|_{k=1}^M)$ to the
zero value of continuous time $x_1=0$
and setting the  amplitudes $\amp_k=1$ we obtain that it
becomes proportional to
\begin{equation}\label{spec}
\Tau^{(0)}_m\sk{0;\{\ff_k,1\}|_{k=1}^M}\sim
\det\left| (-)^{j-1}\sk{{\Om^*_i}^{\;j+m-1}-
{\Mo_i}^{\;j+m-1}}\right|_{i,j=1}^M\;.
\end{equation}
This proves the following
\begin{prop} The rational solitonic tau function
$\Tau^{(0)}_m(0;\{\ff_k,1\}|_{k=1}^M)$ vanishes for values of the
discrete time $m=-M+1,-M+2,\ldots,0$ irrespective the values of the
spectral parameters $\ff_k$.
\end{prop}
This property of the rational tau-function was crucial in \cite{SP1}
in order
to construct  modified $\mathbf{Q}$-operators.

\section{Discussion}

So far the discrete variable $m$ in the previous sections runs over
all integer numbers. It is clear how to apply the results of the previous
section to finite volume system or how to satisfy the periodic
boundary condition with respect to discrete time $m$
\begin{equation}\label{bou-con}
u_m(x_n)=u_{m+M}(x_n)\;,\quad w_m(x_n)=w_{m+M}(x_n)\;.
\end{equation}
One should specialize the spectral variable $\ff$ which uniformize the
curve \r{curve} to the finite set
\begin{equation}\label{set}
\ff_k={2\pi i k\over M}\;, \quad k=1,\ldots,M-1\;.
\end{equation}
The case $k=0$ or $k=M$ are excluded from \r{set} because at this
value of the parameter $\ff$, $\Om^*_\ff=\Mo_\ff$ and
the solitons disappear.
It is clear that in this case there is only finite number ($M$
including a zero soliton case) rational tau-functions.


\section*{Acknowledgment}

The authors are grateful to G.~von Gehlen, S.~Kharchev,
E.~Sklyanin and F.~Smirnov
for useful discussions and comments.

S.P. would also like to thank Max-Planck Instiut f\"ur
Mathematik (Bonn) for support and hospitality and S.S. would thank the
hospitality of MPIM during his short visit to Bonn supported by
Heisenberg-Landau program.

\section*{Appendix: Casoratian technique}

The goal of this Appendix is to demonstrate Casoratian technique
 \cite{ohta} and prove the bilinear relation \r{relation}
with tau-function given by the Casorati determinant \r{oh2}. Let us
fix values of the discrete parameters $m$ and $\nu$. The functions
\r{oh1} satisfy the following dispersion relations
\begin{equation}\label{dispersion}
\begin{array}{rcl}
\ds \frac{\pr\vph{m+\nu}_{m+k}}{\pr x_1}&=&\ds
\sqrt{1+\ka}\;\vph{m+\nu}_{m+k+1}\;,\\[2mm]
\ds \sqrt{1+\ka}\sk{\vph{m+\nu}_{m+k}-\vph{m+\nu-1}_{m+k}}&=&
\vph{m+\nu}_{m+k+1}\;,\quad k\in\ZZ\;.
\end{array}
\end{equation}
Let $\vvph{k}_m$ means the column
\begin{equation}\label{column}
\vvph{\nu}_m=\sk{\begin{array}{c}\vph{m+\nu}_m(x_1;\ff_1,f_1)\\
\vdots\\ \vph{m+\nu}_m(x_1;\ff_M,f_M)
\end{array}}
\end{equation}
and consider the following
identity for $2M\times 2M$ determinant
\begin{equation}\label{iden}
\left|
\begin{array}{ccccccccc}
\vvph{\nu-1}_m&\cdots&\vvph{\nu-1}_{m+M-3}&\vvph{\nu-1}_{m+M-1}&
\vvph{\nu}_{m+M-2}& &0& &\vvph{\nu-1}_{m+M-2}\\
&0& &\vvph{\nu-1}_{m+M-1}& \vvph{\nu}_{m+M-2}&
\vvph{\nu-1}_{m-1}&\cdots&\vvph{\nu-1}_{m+M-3}&\vvph{\nu-1}_{m+M-2}
\end{array}
\right|=0\;.
\end{equation}
The identity is valid since this determinant can be transformed to
$$
\left|
\begin{array}{cccccccccc}
\vvph{\nu-1}_m&\cdots&\vvph{\nu-1}_{m+M-3}&0&
0& -\vvph{\nu-1}_{m-1} & &0& &0 \\
&0& &\vvph{\nu-1}_{m+M-1}& \vvph{\nu}_{m+M-2}&
0&\vvph{\nu-1}_{m}&\cdots&\vvph{\nu-1}_{m+M-3}&\vvph{\nu-1}_{m+M-2}
\end{array}
\right|
$$
by adding and subtracting the columns and rows. Applying the Laplace
expansion to the l.h.s. of identity \r{iden} we obtain the bilinear
relation for $M\times M$ determinants
$$
\begin{array}{rcl}
\left|\vvph{\nu-1}_{m}\cdots\vvph{\nu-1}_{m+M-3}\
\vvph{\nu-1}_{m+M-1}\ \vvph{\nu}_{m+M-2} \right|&\cdot&
\left|\vvph{\nu-1}_{m-1}\ \vvph{\nu-1}_{m}\cdots
\vvph{\nu-1}_{m+M-3}\ \vvph{\nu-1}_{m+M-2} \right|-\\[5mm]
-\left|\vvph{\nu-1}_{m}\cdots\vvph{\nu-1}_{m+M-3}\
\vvph{\nu-1}_{m+M-2}\ \vvph{\nu}_{m+M-2} \right|&\cdot&
\left|\vvph{\nu-1}_{m-1}\ \vvph{\nu-1}_{m}\cdots
\vvph{\nu-1}_{m+M-3}\ \vvph{\nu-1}_{m+M-1} \right|-\\[5mm]
-\left|\vvph{\nu-1}_{m}\cdots\vvph{\nu-1}_{m+M-3}\
\vvph{\nu-1}_{m+M-1}\ \vvph{\nu-1}_{m+M-2} \right|&\cdot&
\left|\vvph{\nu-1}_{m-1}\ \vvph{\nu-1}_{m}\cdots
\vvph{\nu-1}_{m+M-3}\ \vvph{\nu}_{m+M-2} \right|=0\;.
\end{array}
$$
Using now the dispersion relations \r{dispersion}
we can identify
$$
\begin{array}{rcl}
\ds\left|\vvph{\nu-1}_{m-1}\ \vvph{\nu-1}_{m}\cdots
\vvph{\nu-1}_{m+M-3}\ \vvph{\nu-1}_{m+M-2}
\right|&=&\ds\tau^{(\nu-1)}_{m-1}\;,
\\[3mm]
\ds\left|\vvph{\nu-1}_{m}\cdots\vvph{\nu-1}_{m+M-3}\
\vvph{\nu-1}_{m+M-1}\ \vvph{\nu}_{m+M-2} \right|
&=&\ds\frac{1}{1+\ka}\frac{\pr\tau^{(\nu)}_{m}}{\pr x_1}-\tau^{(\nu)}_{m}\;,
\\[3mm]
\ds\left|\vvph{\nu-1}_{m}\cdots\vvph{\nu-1}_{m+M-3}\
\vvph{\nu-1}_{m+M-2}\ \vvph{\nu}_{m+M-2} \right|
&=&\ds\frac{1}{\sqrt{1+\ka}}\tau^{(\nu)}_{m}\;,
\\[3mm]
\ds\left|\vvph{\nu-1}_{m-1}\ \vvph{\nu-1}_{m}\cdots
\vvph{\nu-1}_{m+M-3}\ \vvph{\nu-1}_{m+M-1} \right|
&=&\ds\frac{1}{\sqrt{1+\ka}}\frac{\pr\tau^{(\nu-1)}_{m-1}}{\pr x_1}\;,
\\[3mm]
\ds\left|\vvph{\nu-1}_{m-1}\ \vvph{\nu-1}_{m}\cdots
\vvph{\nu-1}_{m+M-3}\ \vvph{\nu}_{m+M-2} \right|
&=&\ds\tau^{(\nu)}_{m-1}
\end{array}
$$
so the bilinear identity for determinants becomes the bilinear
identity for tau-functions
$$
\sk{\frac{1}{1+\ka}\frac{\pr\tau^{(\nu)}_{m}}{\pr x_1}-\tau^{(\nu)}_{m}}\cdot
\tau^{(\nu-1)}_{m-1}-
\frac{1}{\sqrt{1+\ka}}\tau^{(\nu)}_{m}\cdot
\frac{1}{\sqrt{1+\ka}}\frac{\pr\tau^{(\nu-1)}_{m-1}}{\pr x_1}
+\tau^{(\nu-1)}_{m}
\tau^{(\nu)}_{m-1}=0
$$
which can be written in the form
$$
D_1\sk{\tau^{(\nu-1)}_{m-1}\circ \tau^{(\nu)}_{m}}=
(1+\ka)\sk{\tau^{(\nu-1)}_{m} \tau^{(\nu)}_{m-1}-
\tau^{(\nu-1)}_{m-1}
\tau^{(\nu)}_{m}}\;.
$$
Last equality coincide with \r{relation} at $\nu=0$.

\end{document}